\documentclass[letterpaper, 10 pt, conference]{ieeeconf}  
\usepackage{amsmath,amsfonts}
\include{pythonlisting}
\usepackage{algorithmic}
\usepackage{algorithm}
\usepackage[algo2e,ruled,vlined]{algorithm2e}
\usepackage{array}%\usepackage[caption=false,font=normalsize,labelfont=sf,textfont=sf]{subfig}
\usepackage{textcomp}
\usepackage{stfloats}
\usepackage{url}
\usepackage{verbatim}
\usepackage{graphicx}
\usepackage{cite}
\usepackage{url} 
\usepackage{soul}
\usepackage{multirow}
\usepackage{nicematrix}
\usepackage{subfigure} 
\usepackage{xcolor}
\usepackage{colortbl}

\IEEEoverridecommandlockouts                             
\usepackage[utf8]{inputenc}
\usepackage[T1]{fontenc}

\title{\LARGE \bf
Roadside Units Assisted Localized Automated Vehicle Maneuvering: An Offline Reinforcement Learning Approach
}

\author{Kui Wang$^{1}$, Changyang She$^{2}$, Zongdian Li$^{1}$, Tao Yu$^{1}$, Yonghui Li$^{2}$, and Kei Sakaguchi$^{1}$% <-this % stops a space
\thanks{*This work was supported by the Japan National Institute of Information and Communications Technology (NICT) under JUNO Grant 22404, the Japan Science and Technology Agency (JST) under SPRING Grant JPMJSP2106, and by the Tokyo Tech Academy for Super Smart Society.}% <-this % stops a space
\thanks{$^{1}$K. Wang, Z. Li, T. Yu and K. Sakaguchi are with the Tokyo Institute of Technology, Japan (Email: \{kuiw, lizd, yutao, sakaguchi\}@mobile.ee.titech.ac.jp).}%
\thanks{$^{2}$C. She and Y. Li are with the University of Sydney, Australia (Email: \{changyang.she, yonghui.li\}@sydney.edu.au).}%
}

\begin{document}

\maketitle
\thispagestyle{empty}
\pagestyle{empty}

%%%%%%%%%%%%%%%%%%%%%%%%%%%%%%%%%%%%%%%%%%%%%%%%%%%%%%%%%%%%%%%%%%%%%%%%%%%%%%%%
\begin{abstract}

Traffic intersections present significant challenges for the safe and efficient maneuvering of connected and automated vehicles (CAVs). This research proposes an innovative roadside unit (RSU)-assisted cooperative maneuvering system aimed at enhancing road safety and traveling efficiency at intersections for CAVs. We utilize RSUs for real-time traffic data acquisition and train an offline reinforcement learning (RL) algorithm based on human driving data. Evaluation results obtained from hardware-in-loop autonomous driving simulations show that our approach employing the twin delayed deep deterministic policy gradient and behavior cloning (TD3+BC), achieves performance comparable to state-of-the-art autonomous driving systems in terms of safety measures while significantly enhancing travel efficiency by up to 17.38\% in intersection areas. This paper makes a pivotal contribution to the field of intelligent transportation systems, presenting a breakthrough solution for improving urban traffic flow and safety at intersections.

\end{abstract}

%%%%%%%%%%%%%%%%%%%%%%%%%%%%%%%%%%%%%%%%%%%%%%%%%%%%%%%%%%%%%%%%%%%%%%%%%%%%%%%%
\section{Introduction}
Traffic intersections are one of the most hazardous areas in transportation systems due to frequent blind spots, vulnerable pedestrians, and stop-and-go traffic \cite{yuan2018approach}. A report from the US Federal Highway Administration (FHA) reveals that intersections are hotspots for vehicular accidents, with 40\% of all traffic crashes and 50\% of serious collisions occurring in these areas \cite{trafficaccidents}. Despite the advent and increasing adoption of connected and automated vehicles (CAVs), ensuring their safety at intersections, particularly those without signalization, remains a formidable challenge \cite{qian2019towards}.

Recent advances in vehicle-to-everything (V2X) technologies have given rise to vehicle-road collaboration, which holds promise for enhancing road safety and traffic efficiency \cite{pourjafari2024navi}. V2X technology includes vehicle-to-vehicle (V2V), vehicle-to-infrastructure (V2I), vehicle-to-pedestrian (V2P), and vehicle-to-cloud (V2C) communication \cite{li2023het}. Notably, V2I communication enables real-time data exchange between CAVs and smart roadside infrastructure, laying the groundwork for intelligent transportation ecosystems \cite{suo2023proof}. This synergy between CAVs and roadside infrastructure aids in mitigating the inherent risks associated with traffic intersections by leveraging shared data to optimize driving decisions.

In light of these advancements, there is an increasing focus on integrating roadside infrastructure functionalities, e.g., smart traffic signals and roadside units (RSUs), into the CAV operational frameworks. This integration can be broadly categorized into two key areas: cooperative perception and cooperative driving. Cooperative perception extends the sensory capabilities of vehicles, enabling them to detect potential hazards and traffic conditions beyond their sensing views \cite{caillot2022cpsurvey}. On the other hand, cooperative driving allows CAVs to coordinate their maneuvers in complex environments like traffic intersections and highways, thereby improving overall traffic safety and efficiency.

Intersection-tailored cooperative perception and driving have witnessed significant advancements in recent years. On the one hand, various studies have focused on enhancing CAV operations by using cooperative perception \cite{sakaguchi2021towards, cai2023consensus, meng2023hydro, chang2023bev, xiao2023overcoming, komol2023deep, tan2024dynamic, ren2024interruption, wu20233d}. For example, authors in \cite{chang2023bev} propose a birds-eye-view grid occupancy prediction method to support various applications like driving safety warning and cooperative traffic control. In \cite{xiao2023overcoming}, a perception task-oriented information-sharing network is developed to enable occlusion-free environmental awareness for CAVs. Especially, the work in \cite{komol2023deep, tan2024dynamic} also notices the importance of unique features of different intersections, where intersection-specific strategies are deployed for driver-intended movements prediction and road-to-vehicle visual information sharing, respectively. On the other hand, research in cooperative driving mainly focuses on developing sophisticated algorithms and robust communication protocols that enhance the collaborative decision-making capabilities of CAVs in dynamic and uncertain environments \cite{zhang2021trajectory, yang2021cooperative, xu2023multi, wang2023stop}. For instance, multi-vehicle coordination systems utilize real-time data sharing and predictive modeling to enable vehicles to anticipate the actions of others and adjust their trajectories accordingly, thereby reducing the risk of collisions and improving traffic flow \cite{zhang2021trajectory, yang2021cooperative, xu2023multi}. Advanced intersection management models integrate traffic signal optimization and vehicle trajectory planning, considering factors such as speed, density, and pedestrian movements, to minimize congestion and enhance safety \cite{wang2023stop}.

Most existing research in cooperative perception and cooperative driving focuses on sharing dynamic information like vehicle states and predicted behaviors. However, the integration of static information and inherent road traffic features to develop local strategies can also significantly advance autonomous driving. For example, a human driver commuting daily past an intersection near a kindergarten will quickly learn to pay extra attention and adopt a more careful and conservative driving strategy. How to enable CAVs to gain such knowledge is the core motivation of this research. 

In this paper, we present a novel RSU-assisted cooperative maneuvering system that leverages offline reinforcement learning (RL) to enhance CAV operational strategies at intersections. Our approach uniquely integrates the edge computing capabilities of RSUs \cite{wang2024smdt} with an innovative offline RL framework to create a system that not only learns from historical traffic data but also adapts to the diverse characteristics of different intersections. This methodology enables
the development of highly localized and effective driving strategies that account for unique configurations and dynamic traffic environments of each intersection. Simulation results demonstrate that the proposed system offers a substantial enhancement in traffic safety and efficiency within RSU-installed areas, outperforming conventional autonomous driving
systems. In summary, our key contributions include:

\begin{figure*}[t]
    \centerline{\includegraphics[width=0.98 \textwidth]{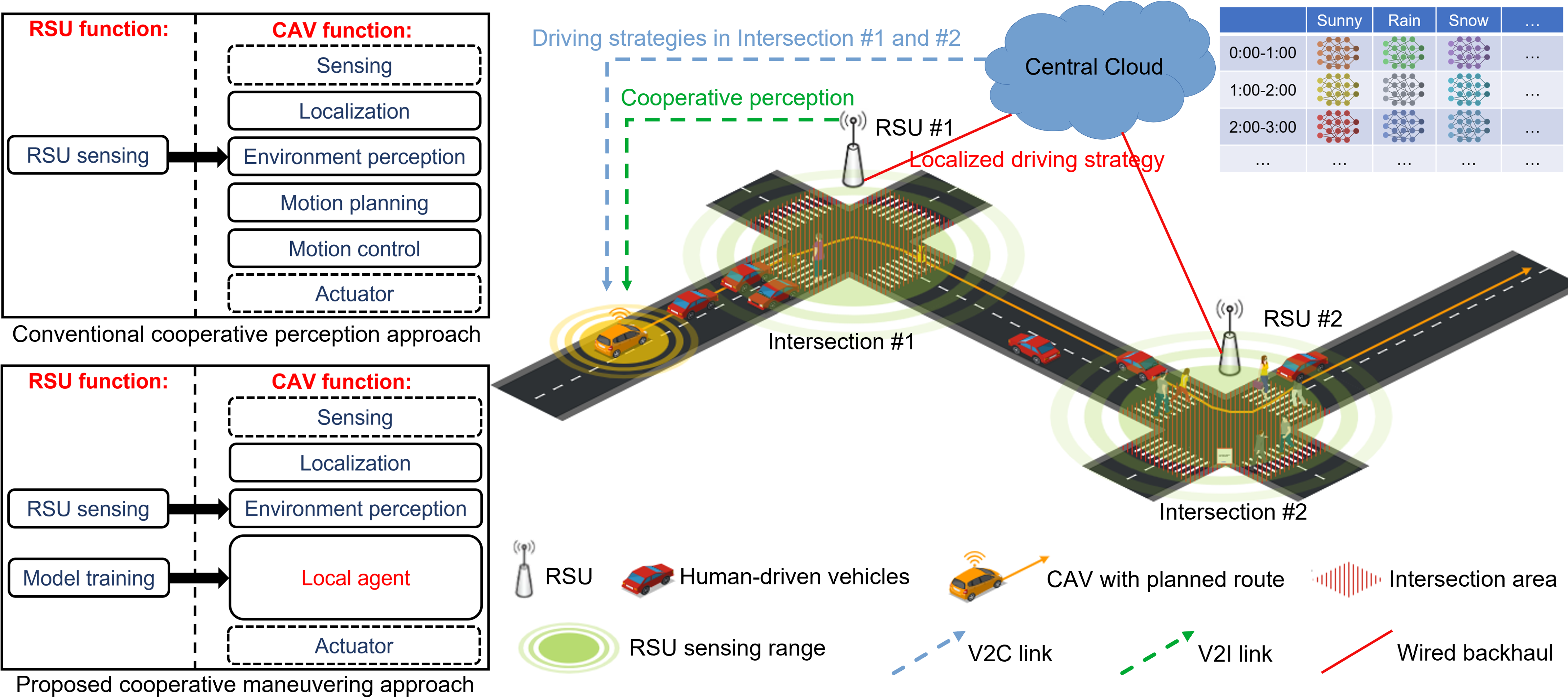}}
    \caption{Conceptual design of RSU-assisted cooperative maneuvering system}
    \label{fig: sysd}
\end{figure*}

\begin{itemize}
    \item \emph{System Design}: We propose a conceptual design of a novel system that uses RSUs to assist CAV maneuvering at intersections by leveraging edge computing and V2X communications for object detection and cooperative perception.
    \item \emph{Intersection-Specific Strategies}: We facilitate localized driving strategies tailored to the unique configurations and dynamic traffic conditions of individual intersections based on the system design.
    \item \emph{Offline RL}: We deploy an innovative offline RL framework that utilizes historical traffic data to optimize driving strategies for CAVs.
    \item \emph{Safety and Efficiency Improvements}: The proposed system demonstrates significant improvements in traffic efficiency and safety in RSU-installed areas, as evidenced by comparative simulations with conventional autonomous driving systems.
\end{itemize}

The remainder of the paper is organized as follows: Section II provides a conceptual architecture of the RSU-assisted cooperative maneuvering system. The application of offline RL for localized motion planning and control is described in Section III. Section IV discusses the evaluation results, and Section V concludes this paper.

\section{RSU-assisted Vehicle Maneuvering System}

We first propose the RSU-assisted cooperative maneuvering system that aims to enhance traffic safety and efficiency for CAVs at non-signalized intersections. In the system, we assume real-time object detection and tracking with the RSU edge computing, high automation of CAV, substantial computational and storage capabilities in cloud computing, as well as connectivity among these smart entities. As RSU can continuously monitor the traffic environment using LiDAR sensors and point cloud-based detection algorithm, the critical role of the RSU in this study is to analyze the driving behaviors of vehicles as they travel through the intersection and to generate a \emph{localized driving strategy} that addresses the unique features of each traffic intersection.

As can be seen from the conceptual system design in Fig.~\ref{fig: sysd}, it is assumed that RSUs are installed at each non-signalized intersection with complex driving conditions or high traffic volume. Such intersections are often fraught with challenges, including the frequent presence of pedestrian crossings and blind spots, which pose potential risks to autonomous vehicles. RSUs keep monitoring traffic flow, continually update the \emph{localized driving strategy}, and synchronize the strategy with a central cloud infrastructure. The cloud acts as a dynamic repository for a diverse array of RSU-trained models, each tailored to specific environmental conditions, such as varying times of day and weather patterns. This intricate setup allows the cloud to perform real-time assessments of current conditions and select the most appropriate model to guide CAVs effectively. For example, as depicted in Fig.\ref{fig: sysd}, a CAV scheduled to traverse intersections \#1 and \#2 would receive the \emph{localized driving strategies} that match current time and weather from the central cloud via V2C communication. Upon the CAV entering the V2I communication range with RSU, the RSU serves a role in providing cooperative perception, expanding the horizon of the onboard sensing view. This enhanced perception aids the CAV in gaining a comprehensive understanding of the traffic environment and in identifying potential hazards concealed within blind spots.

Our system design capitalizes on the edge computing capabilities of RSUs and CAVs by allocating specific functions to each, considering the edge's nature to facilitate real-time operations for delay-sensitive tasks \cite{wang2023dtad}. Hence, cooperative environmental perception is managed by RSUs, while vehicle maneuvering is delegated to CAVs, mitigating latency in critical decision-making processes. To avoid the long communication delay caused by large data rates of raw sensor data, edge computing on RSUs is employed to process the real-time LiDAR point cloud into object-level environmental perceptions, then share the real-time perception to the passing by CAVs through V2I link. Furthermore, we employ the cloud for disseminating \emph{localized driving strategies} to CAVs, leveraging the cloud's expansive storage capabilities to house a comprehensive library of pre-trained models for various intersections in various conditions. 

\section{Offline RL Based Vehicle Maneuvering}
In this section, we propose an offline RL approach for training an agent for autonomous vehicle motion planning policy using a dataset obtained from real-world RSU. In particular, we introduce the formulation of a partially observed Markov decision process (POMDP) and explain the mechanism for achieving vehicle motion control.

\subsection{Intersection Environment and Data Processing}

In our study, the intersection environment is monitored by RSUs, capable of detecting, tracking, and identifying all traffic participants within their sensor range, which covers the entire intersection area. We define $\mathcal{P}(t) = \{p_1, p_2, \ldots, p_N\}$ as the set representing all detected traffic participants within the intersection at time step $t$. For each participant $p_k$ in $\mathcal{P}(t)$, the state is defined as follows:

\[s_k(t) = [type_k, x_k, y_k, \theta_k, v_k, v_{lon,k}, v_{lat,k}]\]

\noindent where $type_k$ shows the type of the detected object, which could be a pedestrian, vehicle, or cyclist. $x_k$ and $y_k$ denote the absolute position coordinates of the center point of $p_k$. The heading angle, or yaw angle, is represented by $\theta_k$. The velocities $v_k$, $v_{lon,k}$, and $v_{lat,k}$ denote velocity, longitudinal velocity, and lateral velocity of $p_k$, respectively. The set $\mathcal{S}(t) = \{s_1, s_2, \ldots, s_N\}$ indicates the state space at time step $t$ describing the states of all traffic participants detected by RSU.

\subsection{Formulation of POMDP}

Offline RL operates on the principle of learning optimal policies from a dataset of previous experiences, without the need for further interaction with the environment \cite{levine2020offline}. In the context of vehicle maneuvering at intersections, this approach enables the utilization of historical traffic data to inform decision-making processes. 

Offline RL in the context of a POMDP is formally described by a tuple $M = (\mathcal{S}, \mathcal{A}, \mathcal{O}, \mathcal{T}, \mathcal{Z}, \mathcal{R}, \gamma)$. $\mathcal{S}$, $\mathcal{A}$, and $\mathcal{O}$ represent the sets of states, available actions, and observations that the agent can perceive in the environment. $\mathcal{T}(s'| a, s)$ defines the probability of transitioning to state $s'$ from state $s$ after taking action $a$. $\mathcal{Z}(o| s)$ is an emission
function defining the distribution from $s$ to $o$. $\mathcal{R}$ is the reward function and $\gamma \in (0 ,1]$ is the discount factor.

\textbf{Observation:} At a specific time step $t$. We denote the vehicle traversing the intersection as $p_{j} \in \mathcal{P}(t)$, where $type_j=\textrm{`Vehicle'}$. Given the dynamic nature of traffic environments, where the total number of objects $N$ fluctuates, it is impractical to include every detected object within the observation space. Instead, we focus on individual vehicles and their surroundings.

\begin{figure}[t]
    \centerline{\includegraphics[width=0.48\textwidth]{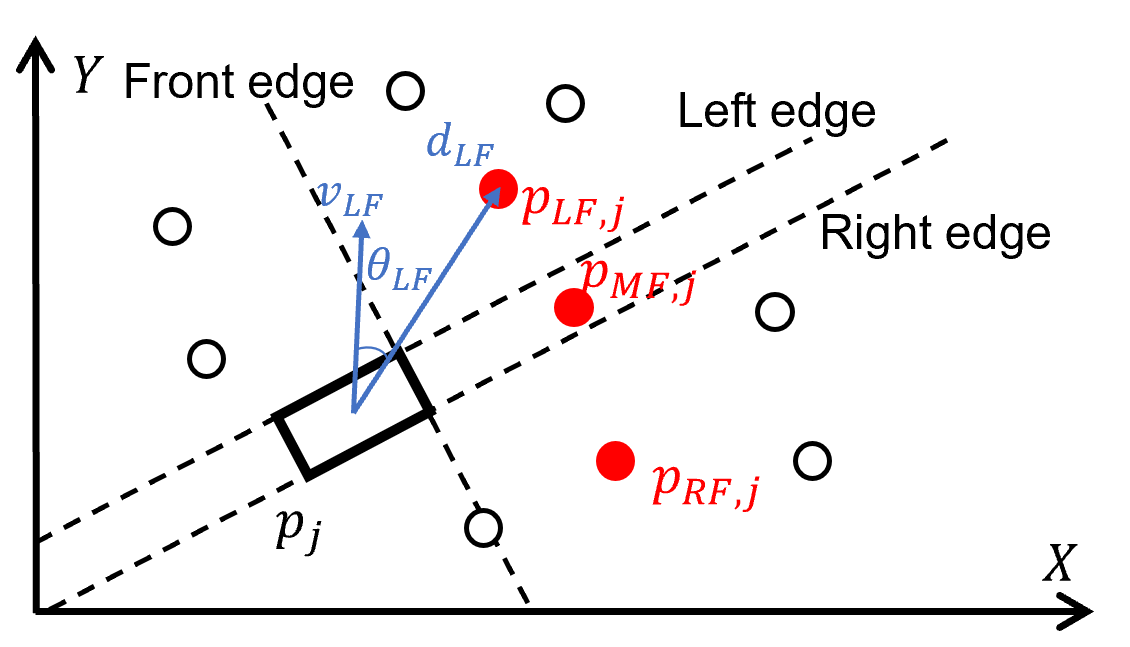}}
    \caption{Illustration of observation space and relative velocity}
    \label{fig: obs}
\end{figure}

The obstacles in the surrounding area, particularly along the vehicle's heading direction, are critical for road safety. We classify the surrounding area of the vehicle $p_{j}$ into ``left-front'' (LF), ``middle-front'' (MF), ``right-front'' (RF), and ``back'' areas, based on the vehicle’s frontal and lateral boundaries.

In the LF, MF, and RF areas, the closest objects to the vehicle $p_{j}$ are denoted by $p_{LF, j}$, $p_{MF, j}$, and $p_{RF, j}$, respectively, as shown in Fig.~\ref{fig: obs}. These nearest objects represent the most immediate and potentially impactful elements on the vehicle’s driving behavior, especially in terms of collision avoidance and speed adjustment.

Thus, at any given time $t$, the observation $o(t) \in O$ is defined as follows:

\[o(t) = [s_{j}, s_{LF, j}, s_{MF, j}, s_{RF, j}]\]

\noindent where $s_{j}$ is the state of the vehicle $p_{j}$, and $s_{LF, j}$, $s_{MF, j}$, and $s_{RF, j}$ represent the states of the nearest objects within the LF, MF, and RF areas, respectively. 

\textbf{Action:} To effectively accommodate the diverse mechanical constraints of different vehicles, such as baseline length, steering limits, and acceleration capabilities, the action $a$ should be designed considering three requirements: \textit{(i)} capabilities to reflect strategic driving decisions in the traffic environment, \textit{(ii)} it can be derived from the vehicle's state $s_{j}$, and \textit{(iii)} can be translated into the control space for different automated vehicles. Specifically, for the vehicle $p_{j}$ at time $t$, the action $a(t)$ is designed as follows:

\begin{equation}
    a(t) = [v_{x, j}, v_{y, j}, \dot{\psi}_{j}]
\end{equation}

\noindent where $v_{x}$ and $v_{y}$ represent the vehicle's velocity components along the $X$-axis and $Y$-axis, respectively, and $\dot{\psi}$ denotes the yaw rate. Here, the origin of the coordinate system is the location of RSU. As such, the actions of all vehicles are represented in the same coordinate system. Thus the projection of velocity $v_{x}$ and $v_{y}$ onto this coordinate system becomes an insightful indicator of the vehicle's driving strategy.

\textbf{Reward:} To effectively evaluate the performance of driving strategies within our framework, the reward function is constructed as a linear combination of three critical components: road safety, traveling efficiency, and deviation from the road centerline. 

\[r = \alpha_1 r_{\textrm{safety}} + \alpha_2 r_{\textrm{effi}} + \alpha_3 r_{\textrm{dev}}\]

\noindent where $\alpha_1$,  $\alpha_2$, and  $\alpha_3$ serve as weighting coefficients.

The safety component $r_{\textrm{safety}}$ is defined with the time-to-collision (TTC) between the vehicle and the three closest objects. For example, considering the closest object on the LF side of the vehicle $p_{LF, j}$, we use $d_{LF, j}$ and $v_{LF, j}$ to represent the relative distance and relative velocity between the $p_{j}$ and $p_{LF, j}$, and $\theta_{LF, j}$ to define the angle between $d_{LF, j}$ and $v_{LF, j}$, as shown in Fig.~\ref{fig: obs}. In a short time interval, it can be assumed that the vehicle and the object do not change their velocity, then the TTC can be expressed as:

\begin{equation}
    TTC_{LF, j} = 
    \begin{cases}
        \frac{d_{LF, j}}{v_{LF, j} \cos{\theta_{LF, j}}} & \text{if } \theta_{LF ,j} \in [0, \frac{\pi}{2})  \\
        1 & \text{if } \theta_{LF, j} \in [\frac{\pi}{2}, \pi]
    \end{cases}
\end{equation}

Similar to the calculation of $TTC_{LF, j}$ in equation (2), we can obtain $TTC_{MF, j}$ and $TTC_{RF, j}$ for the closest objects on the MF and RF sides. $r_{\textrm{safety}}$ is then determined by:

\[
r_{\textrm{safety}} = \sum_{i = LF, MF, RF} \frac{\min [TTC_{i ,j}, TTC_{\textrm{thre}}]}{TTC_{\textrm{thre}}}
\]

\noindent where $TTC_{\textrm{thre}}$ is a predefined threshold that serves as a boundary to evaluate the immediacy of potential collisions. 

The efficiency reward component $r_{\textrm{effi}}$ is defined as the optimal travel speeds normalized by the maximum speed,

\[
r_{\textrm{effi}} = 
    \begin{cases}
        \frac{v_{j}}{v_{\textrm{max}}} & \text{if } v_{j} \leq v_{\textrm{max}}  \\
        - \frac{v_{j} - v_{\textrm{max}}}{v_{\textrm{max}}} & \text{if } v_{j} > v_{\textrm{max}}, 
    \end{cases}
\]

\noindent where $v_{\textrm{max}}$ represents the speed limit. When $0 \leq v_{j} \leq v_{\textrm{max}}$, $r_{\textrm{effi}}$ is positive to encourage velocity maintenance near the speed limit. When $v_{j} > v_{\textrm{max}}$, $r_{\textrm{effi}}$ becomes negative for the penalty of overspeeding.

Lastly, the reward for minimizing deviation from the lane's centerline $r_{\textrm{dev}}$, is designed to incentivize precise lane-keeping by penalizing deviations from the centerline:

\[
r_{\textrm{dev}} = 1-\frac{2\Delta}{w}
\]

\noindent where $\Delta$ and $w$ represent the lateral displacement from the centerline and the average lane width.

\subsection{Kinematic Model for Vehicle Maneuvering}

\begin{figure}[t]
    \centerline{\includegraphics[width=0.48\textwidth]{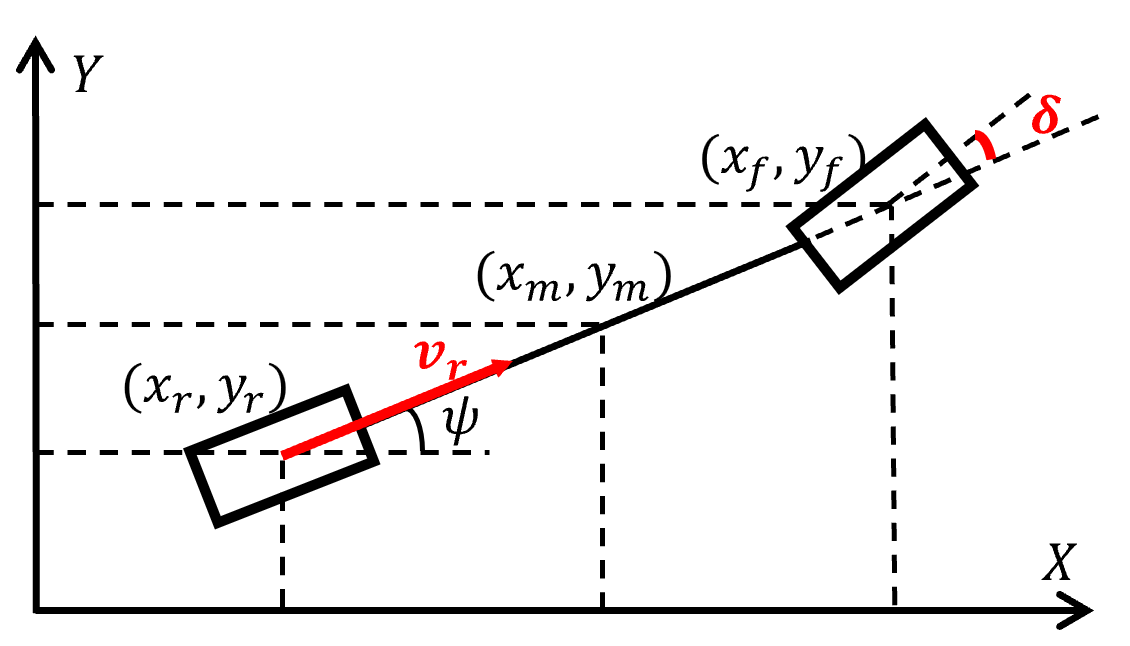}}
    \caption{Illustration of kinematic bicycle model}
    \label{fig: kine}
\end{figure}

To achieve vehicle motion control with the action space in equation (1), we introduce a kinematic bicycle model to simplify car-like vehicle dynamics, as shown in Fig.~\ref{fig: kine}. The controlled vehicle is assumed to have front-wheel steering and rear-wheel drive. We use $(x_{\textrm{f}}, y_{\textrm{f}})$, $(x_{\textrm{r}}, y_{\textrm{r}})$ and $(x_{\textrm{m}}, y_{\textrm{m}})$ to represent the coordinates of centers of the front axle, rear axle, and the center point of the baseline. $v_{\textrm{f}}$ and $v_{\textrm{r}}$ are the velocities of front and rear wheels. The yaw angle and steering angle are represented by $\psi$ and $\delta$. $l$ is the length of the vehicle baseline. At the center of the rear axle, the velocity $v_{\textrm{r}}$ can be expressed with its projections on the $X$-axis and $Y$-axis:

\begin{equation}
v_{\textrm{r}} = \dot{x}_{\textrm{r}}\cos{\psi} + \dot{y}_{\textrm{r}}\sin{\psi}
\end{equation}

The position of the middle point of the vehicle baseline is:

\begin{equation}
\left\{
\begin{aligned}
x_{\textrm{m}} &= x_{\textrm{r}} + \frac{l}{2}\cos{\psi} \\
y_{\textrm{m}} &= y_{\textrm{r}} + \frac{l}{2}\sin{\psi}
\end{aligned}
\right.
\end{equation}

The differential form of Equation (4) is:

\begin{equation}
\left\{
\begin{aligned}
\dot{x}_{\textrm{m}} &= \dot{x}_{\textrm{r}} - \frac{l\dot{\psi}}{2}\sin{\psi} \\
\dot{y}_{\textrm{m}} &= \dot{y}_{\textrm{r}} + \frac{l\dot{\psi}}{2}\cos{\psi}
\end{aligned}
\right.
\end{equation}

Here, we note that $\dot{x}_{\textrm{m}}$ and $\dot{y}_{\textrm{m}}$ correspond to the $v_x$ and $v_y$ in the action space. Based on equations (3) and (5), the rear tire velocity can be expressed with the $\dot{x}_{\textrm{m}}$ and $\dot{y}_{\textrm{m}}$:

\begin{equation}
v_{\textrm{r}} = \dot{x}_{\textrm{m}}\cos{\psi} + \dot{y}_{\textrm{m}}\sin{\psi}
\end{equation}

At the center of the front and rear axles, the kinematic constraints are:

\begin{equation}
\left\{
\begin{aligned}
&\dot{x_{\textrm{f}}}\sin{(\psi + \delta)} - \dot{y_{\textrm{f}}}\cos{(\psi + \delta)} = 0\\
&\dot{x_{\textrm{r}}}\sin{\psi} - \dot{y_{\textrm{r}}}\cos{\psi} = 0
\end{aligned}
\right.
\end{equation}

Based on equations (3) and (7), we can obtain:

\begin{equation}
\left\{
\begin{aligned}
\dot{x}_{\textrm{r}} &= v_{\textrm{r}}\cos{\psi} \\
\dot{y}_{\textrm{r}} &= v_{\textrm{r}}\sin{\psi}
\end{aligned}
\right.
\end{equation}

According to the relative positions of the front and rear tires, the position of the front tire is:

\begin{equation}
\left\{
\begin{aligned}
x_{\textrm{f}} &= x_{\textrm{r}} + l\cos{\psi}\\
y_{\textrm{f}} &= y_{\textrm{r}} + l\sin{\psi}
\end{aligned}
\right.
\end{equation}

Based on equations (7)-(9), the yaw rate of the vehicle can be expressed as:

\begin{equation}
\dot{\psi} = \frac{v_{\textrm{r}}}{l}\tan{\delta}
\end{equation}

For the autonomous driving motion control, $(v_{\textrm{r}}, \delta)$ is regarded as the control space. According to equations (6) and (10), we translate the action space to the control space, and thus facilitating motion control for automated vehicles.

\section{Evaluation on Hardware-in-loop Autonomous Driving Simulation}
In this section, we first introduce some popular offline RL algorithms and compare their performance on our dataset, i.e., RSU-based object detection. Then we evaluate the safety and efficiency of our approach.

\subsection{Offline RL Algorithms and Performances}

In this research, we benchmark the following existing algorithms:

\subsubsection{Twin Delayed DDPG (TD3) \cite{fujimoto2018addressing}} It enhances the DDPG algorithm by addressing its overestimation bias. It introduces two key modifications: using twin Q-networks to minimize value overestimation and delaying policy updates to stabilize the learning process.

\subsubsection{Behaviour Cloning (BC) \cite{torabi2018behavioral}} It is a straightforward effective machine learning technique where an agent learns to mimic actions from the environment. By directly mapping observed states to actions, BC simplifies the learning process, making it an accessible approach for initiating complex behaviors in autonomous systems.

\subsubsection{TD3+BC \cite{fujimoto2021minimalist}} It combines the stability and efficiency of TD3 with the simplicity of BC. This algorithm leverages TD3's control over estimation bias and BC's proficiency in imitating behaviors, offering a balanced approach for solving challenging RL tasks with improved robustness.

\begin{figure}[t]
    \centerline{\includegraphics[width=0.48\textwidth]{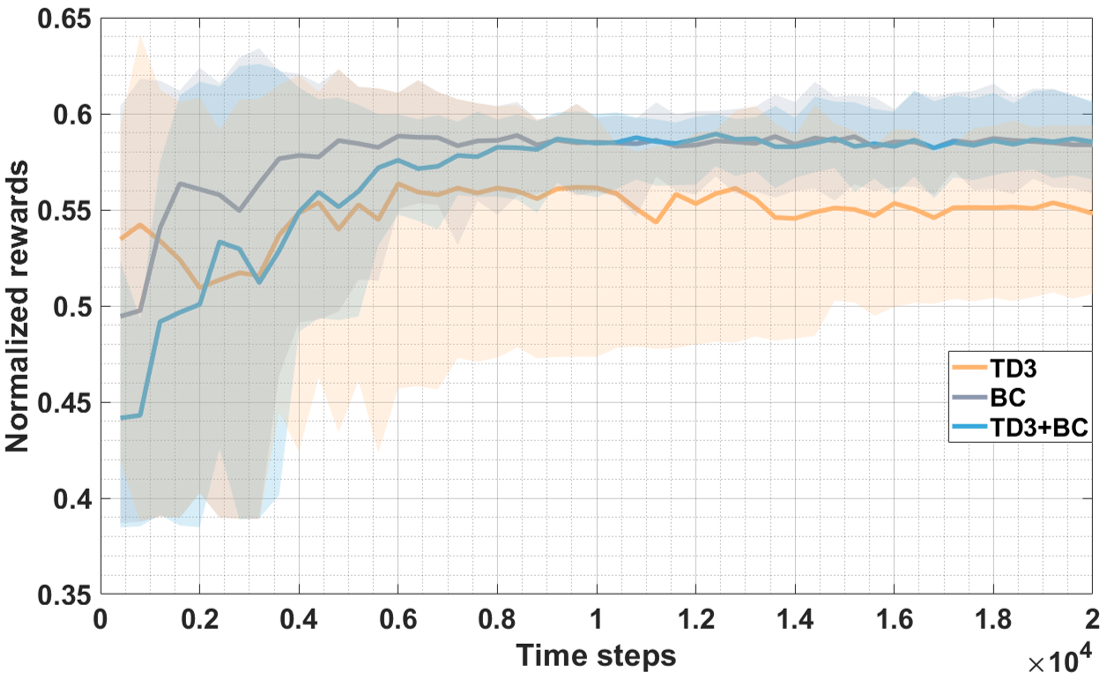}}
    \caption{Performance of offline RL algorithms with 10 random seeds}
    \label{fig: reward}
\end{figure}

We provide the training results of normalized rewards in Fig.~\ref{fig: reward}. It can be seen that the solid lines and dimmed areas represent the mean and confidence interval over 10 random seeds, respectively. The maximum time step to run our customized environment is set to 20,000, with evaluations conducted every 400 time steps. Remarkably, the TD3, BC, and TD3+BC algorithms all achieve convergence, demonstrating the effectiveness of our dataset and the rationality of our POMDP formulation. The convergence results for both the BC and TD3+BC algorithms are around 0.58, surpassing that of the TD3 algorithm. Therefore, in subsequent simulation-based evaluations, we further test and assess the agents trained with the TD3+BC and BC algorithms.

\subsection{Setup in Hardware-in-Loop Simulation}

\begin{figure}[t]
    \centerline{\includegraphics[width=0.48\textwidth]{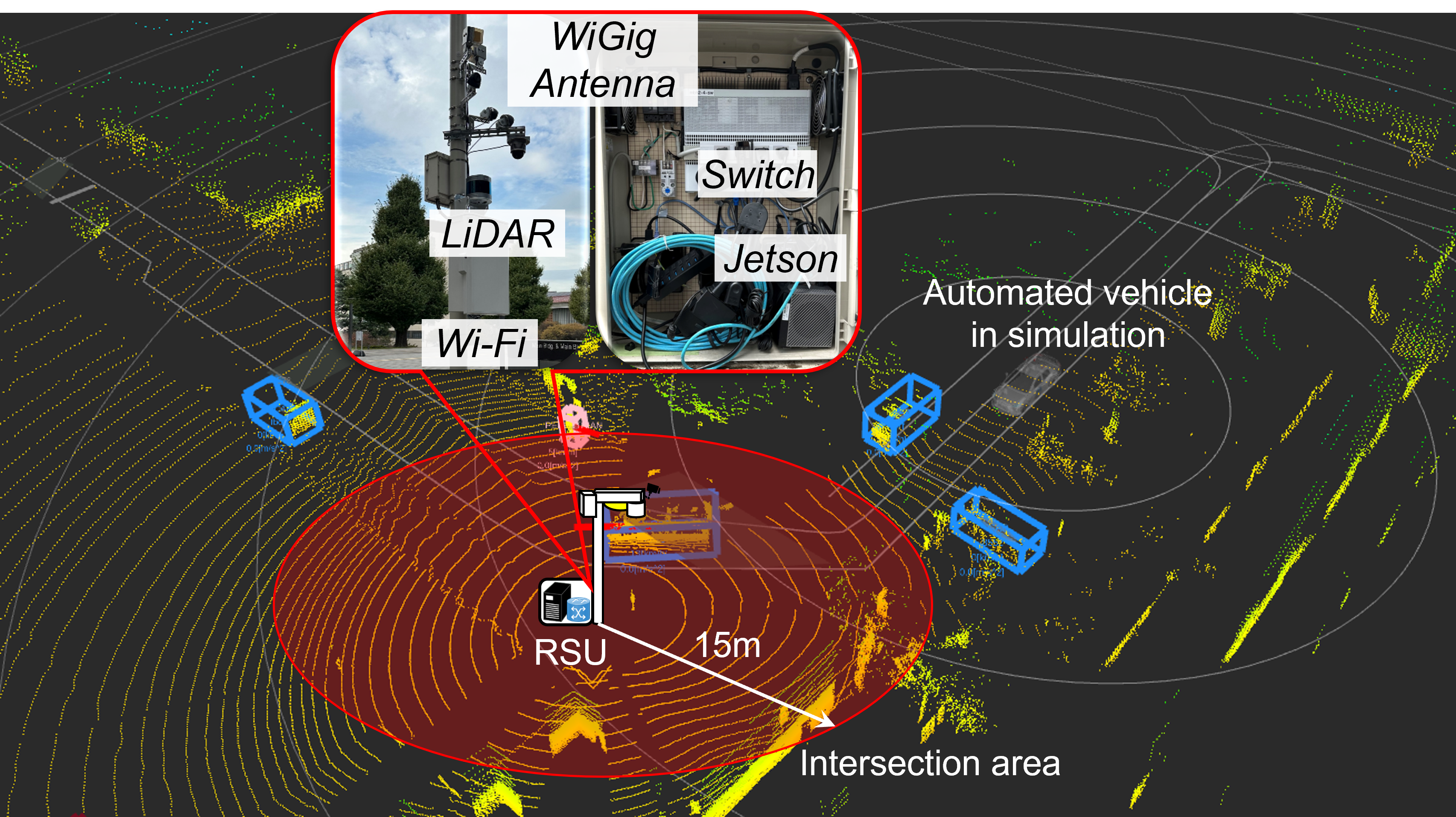}}
    \caption{Simulation environment in Rviz}
    \label{fig: sim}
\end{figure}

The testing is conducted within a simulation environment provided by Autoware \cite{Autoware}, in which we launch object detection based on the real-time point clouds from RSU LiDAR. Then a simulated automated vehicle seeks to transverse the intersection area, avoiding the detected objects shared by RSU. Fig.~\ref{fig: sim} shows an example of initial state of the simulation in a visualization tool, i.e., Rviz. 

In the experiment, we compared the performance of three agents: \textit{(i)} agent trained with TD3+BC algorithm, named TD3+BC agent, \textit{(ii)} agent trained with BC algorithm, named BC agent, and \textit{(iii)} an inherent driving agent of Autoware, named Autoware agent. In the driving process, the automated vehicle always adopts Autoware agent outside the intersection area. When it enters the intersection, we employ the TD3+BC agent, BC agent, and Autoware agent, respectively, to facilitate the autonomous vehicle's navigation through intersections. Additionally, based on the monitored traffic conditions, we categorize the traffic flow into three types: low density (the number of objects within the intersection area is less than 3), middle density (ranging from 3 to 6), and high density (larger than 6). We study the performance of different agents under varying traffic densities. The experiment of each agent in each traffic density is conducted 5 times.

\subsection{Testing Results and Discussions}

\begin{figure*}[t]
\centering
\subfigure[]{
\begin{minipage}[b]{0.314\textwidth}
\includegraphics[width=1\textwidth]{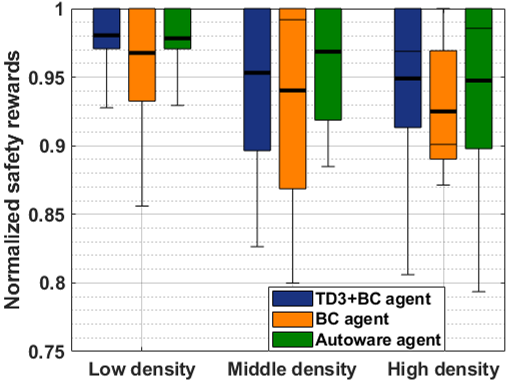} 
\end{minipage}
}
\subfigure[]{
\begin{minipage}[b]{0.314\textwidth}
\includegraphics[width=1\textwidth]{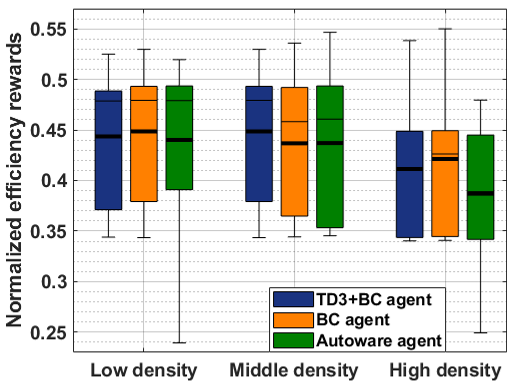} 
\end{minipage}
}
\subfigure[]{
\begin{minipage}[b]{0.314\textwidth}
\includegraphics[width=1\textwidth]{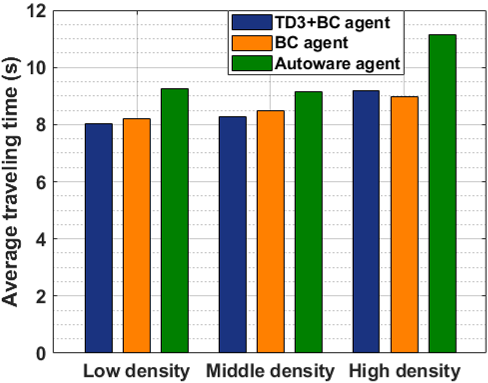} 
\end{minipage}
}
\caption{Hardware-in-loop simulation results. (a) Normalized safety rewards of TD3+BC agent, BC agent, and Autoware agent in different traffic densities, (b) Normalized efficiency rewards, (c) Average traveling time.}
\label{fig: result}
\end{figure*}

Testing results are shown in Fig.~\ref{fig: result}, where box plots are employed to visually represent the distribution of safety and efficiency rewards attained by various agents during transit, and the mean value is delineated using a bold solid line. We also utilize a bar graph to present the average traveling time within intersection areas, thereby offering a more intuitive representation of performance in terms of transit efficiency. 

From the perspective of safety, due to the Autoware agent adopting a relatively conservative and safety-oriented driving strategy, which refrains from actively overtaking obstacles and instead opts to wait for their leaving, the results depicted in Fig.~\ref{fig: result}(a) indicate that the trained agents cannot exhibit safety improvements. Among them, the TD3+BC agent demonstrated better safety measures, closely approximating the Autoware agent's safety rewards across various traffic densities. However, the BC agent's safety rewards were comparatively lower because the BC agent primarily emulates human driving behaviors and strategies, which often entail lower safety levels. Human drivers, due to a myriad of factors such as limited visibility at intersections and the inherent unpredictability of human behavior, tend to exhibit a lower degree of safety during the driving process. 

Regarding efficiency, Fig.~\ref{fig: result}(b) illustrates that the trained agents exhibit considerable enhancements in performance compared to the Autoware agent, especially under conditions of high traffic density. This improvement is further corroborated by Fig.~\ref{fig: result}(c), which displays the average transit times, providing empirical evidence of the enhanced efficiency. In comparison to the Autoware agent, the TD3+BC agent is capable of reducing transit times by up to 17.38\%, while the BC agent achieves a maximum reduction in transit time of 19.43\%. These findings underscore the efficacy of the system and offline RL approach we proposed, demonstrating its capability to enhance traffic efficiency while ensuring safety, particularly in high-traffic scenarios.

\section{Conclusion and Future Works}
In this paper, we have designed an RSU-assisted cooperative maneuvering system to enhance traffic safety and efficiency for CAVs at intersections. Our system utilizes RSUs to capture real-world traffic information, adopts offline RL to study human-driven vehicles' driving behaviors, and then provides local motion control strategies for automated vehicles. Simulation results show that the agent trained with the TD3+BC algorithm can ensure road safety and decrease traveling time in the intersection area by up to 17.38\%, compared to conventional autonomous driving software systems. Our future work includes real-world implementation and testing of the proposed system, with the consideration of communication, scalability, and robustness issues.

\bibliographystyle{IEEEtran.bst}
\bibliography{bibliography}

\end{document}